\documentclass[twocolumn]{article}
\pdfoutput=1
\usepackage[utf8]{inputenc}
\usepackage[margin=0.75in]{geometry} 
\usepackage{graphicx}               	
\usepackage{soul, xcolor}            	
\usepackage{subcaption}              	

\title{ From plague to coronavirus: \\ On the value of ship traffic data for epidemic modeling}
\author{Katherine Hoffmann Pham\textsuperscript{\rm 1}\textsuperscript{,\rm 2} and  Miguel Luengo-Oroz\textsuperscript{\rm 1}  \medskip \\
\textsuperscript{\rm 1}United Nations Global Pulse, 
\textsuperscript{\rm 2}NYU Stern School of Business \smallskip \\
katherine@unglobalpulse.org, miguel@unglobalpulse.org
}
\date{}

\begin{document}
\maketitle

\abstract{In addition to moving people and goods, ships can spread disease. Ship traffic may complement air traffic as a source of import risk, and cruise ships -- with large passenger volumes and multiple stops -- are potential hotspots, in particular for diseases with long incubation periods. Vessel trajectory data from ship Automatic Identification Systems (AIS) is available online and it is possible to extract and analyze this data. We illustrate this in the case of the current coronavirus epidemic, in which hundreds of infected individuals have traveled in ships captured in the AIS dataset. This real time and historical data should be included in epidemiological models of disease to inform the corresponding operational response. }

\section{Introduction}

Faced with a global epidemic, there is a need to quickly estimate potential routes for disease transmission in order to prepare a public health response. In recent years, the growing availability of big data sources has facilitated the modeling of these potential routes. For example, previous studies have used mobile phone records, commuting flow data, and airline flight matrices to identify possible disease hotspots, predict contagion patterns, and/or estimate import risk \cite{broeck_gleamviz_2011, wesolowski_quantifying_2012}. In this research, we highlight another dataset -- Automatic Identification System (AIS) data containing historical and real-time ship trajectories -- as a complementary source of information on possible transmission routes.

AIS data forms a global database of maritime traffic. All passenger ships, international ships over 300 tons, and cargo ships over 500 tons must be equipped with an AIS transceiver, which reports dynamic information about the ship’s position -- such as its location, speed, and course over ground -- as well as static information such as the ship’s identifiers, vessel type, and flag \cite{raymarine_ais_2016}. This data is collected by a network of land- and satellite-based receivers and subsequently aggregated by third-party providers, which redistribute the data and/or render it for viewing in a graphical user interface. AIS data has been used in a diverse range of applications, including monitoring for port security, detecting fishing behavior, and studying search and rescue operations involving migrant and refugee boats \cite{rhodes_maritime_2005, natale_mapping_2015, hoffmann_pham_data_2018}. 

Currently, a standard approach to calculating disease import risk at the international level is epidemic modeling using flight network data. However, vessels can also be global carriers for infected individuals and disease vectors. For example, the ``Black Death’’ plague outbreak in 1347, which killed an estimated 25 million Europeans, is believed to have been spread throughout Europe by merchant ships, possibly by rats on board and their parasites \cite{encyclopaedia_britannica_black_2019}. Similarly, the travel of infected civilians and soldiers along shipping routes is credited with helping to spread the catastrophic 1918 Spanish flu epidemic, which killed an estimated 20 - 40 million people worldwide \cite{billings_1918_2005}. 

Although passenger ships have largely been replaced by airlines in the modern day, ships still represent a large amount of traffic in island nations, and continue to play an important role in international trade. Although ship traffic patterns have transmission potentials that are distinct from airline traffic, they are not typically taken into account in current epidemic models.  Below, we discuss two case studies in which shipping traffic is relevant to the study of disease transmission. 

\section{Analysis}

\begin{figure*}
\centering    
    \begin{subfigure}[t]{0.47\textwidth}
        \centerline{\includegraphics[height=2.75in]{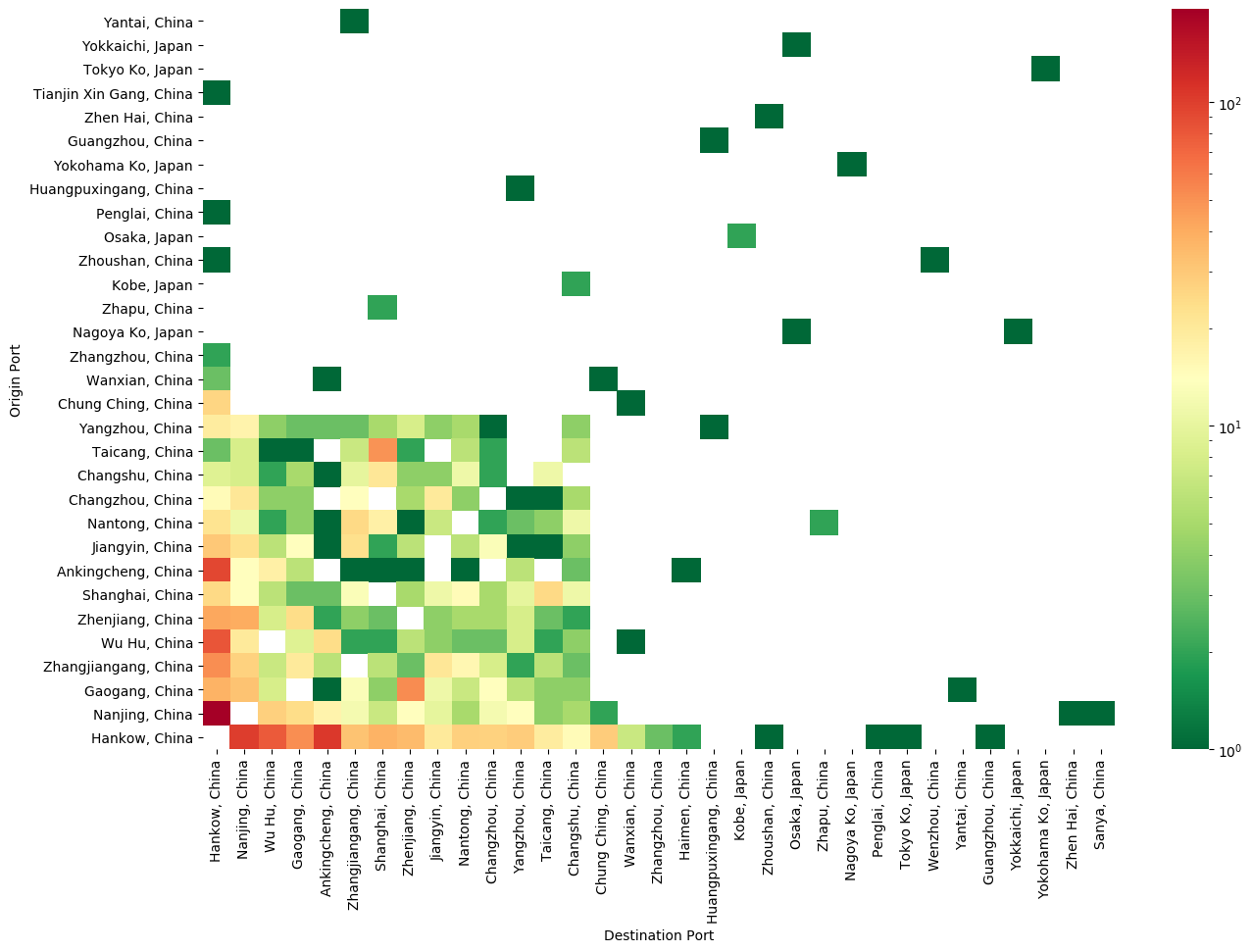}}
        \caption{Origin-destination matrix for ships visiting Wuhan from the 8th to 21st January 2020. \textit{Data source: UN Global Platform AIS Database.}}
        \label{fig:wuhan_od}
    \end{subfigure}
    ~~
    \begin{subfigure}[t]{0.47\textwidth}
        \centerline{\includegraphics[height=2.75in]{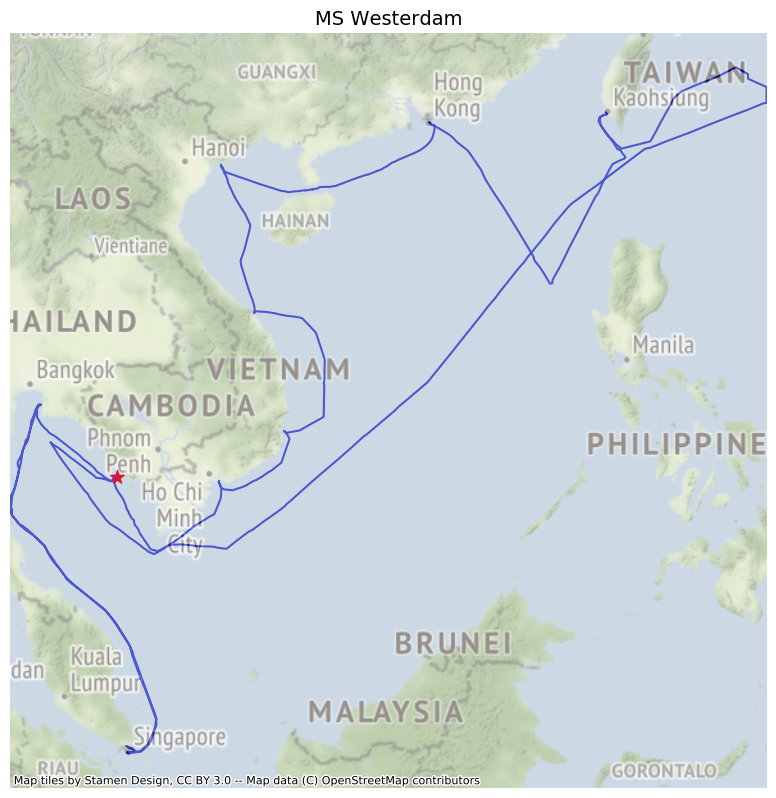}}
        \caption{The tracks of the cruise ship \textit{MS Westerdam} from 1st January to 19th February 2020. \textit{Data source: UN Global Platform AIS Database. Basemap source: Stamen Design and OpenStreetMap contributors.}}
        \label{fig:cruises}
    \end{subfigure}
    \caption{Sample applications of AIS data }
\end{figure*}

\subsection{The 2017 Madagascar plague outbreak}
In Fall 2017 Madagascar was struck by a plague epidemic, affecting an estimated 2,348 individuals and resulting in 202 deaths \cite{world_health_organization_plague_2017}. Since Madagascar is an island nation, we examined travel patterns in the AIS data to estimate possible transmission routes through commercial and cruise ships. Our methodology consisted of gathering information from ports and cruise ships and constructing an origin-destination matrix that captured all trips between ports inside Madagascar ($\sim$ 40\%
 of observed trips) and outside Madagascar ($\sim$ 60\%). 

From 26 September to 26 October 2017 we identified 126 vessel departures from the main ports in Madagascar, including cargo vessels, tankers, tugs, and fishing vessels. Interestingly from an epidemic risk perspective, long-range trips included destinations as diverse as Mayotte, Mauritius, Reunion, Mozambique, France, and China. A further analysis of cruise traffic specifically found 6 cruises that were scheduled to arrive in Madagascar over the course of November 2017. Generally, these cruises were scheduled to stop in more than one port, and the number of passengers and crew members on board was over 1,200 on average. These cruises therefore represent high-volume points of human contact which in this case would have subsequently visited South Africa, Mauritius, Mozambique, Reunion, Seychelles, and Italy.

\subsection{The 2020 coronavirus epidemic}
 Given the long incubation period of the disease and the possibility of mild, asymptomatic cases, maritime traffic has played a significant role in the current spread of the coronavirus epidemic. In particular, experience has shown that large cruise ships can become important hotspots. Below are three scenarios in which AIS information could be used to support further investigations and risk modeling.

\begin{itemize}

    \item Wuhan, the city at the center of the outbreak, is located at the junction of the Yangtze and Han rivers. Therefore, shipping traffic is one pathway by which disease might spread out of the city. Using AIS data from Exact Earth, we obtained all traffic from ships that passed within 50km of Wuhan with a speed of less than 2 knots for the period between 8 - 21 January 2020. All told, we identified 1,107 ships connecting 34 different ports across China and Japan (see Figure \ref{fig:wuhan_od}).  

    \item Spain's first case of the coronavirus was identified in the Canary Islands, in a traveler who arrived in La Gomera by ferry from
    Tenerife \cite{paniagua_en_2020, reuters_update_2020}.  In addition to the islands' air traffic connections, AIS data analysis for 8 - 21 January 2020 showed that ships passing by the Canary Islands were associated with ports in 56 different countries, including frequent travel to peninsular Spain, Morocco, Senegal, Gibraltar, the UK, the Netherlands, and Portugal.

    \item The \textit{Diamond Princess} cruise ship was quarantined off the coast of Japan with approximately 3,700 individuals on board, and 621 people on the ship were ultimately infected \cite{bbc_news_growing_2020}. A second cruise ship, the \textit{MS Westerdam}, was repeatedly denied entry at port for fear that there were infected individuals among the 2,257 people onboard, and a passenger tested positive for coronavirus only after the ship docked in Cambodia, leading to fears that other infected but asymptomatic individuals could spread new cases (although the passenger has since tested negative) \cite{bbc_news_how_2020, ducharme_american_2020}. The locations of both ships are available in real time and their travel paths can be reconstructed using historical AIS traces (see Figure \ref{fig:cruises}). Similar data on positions, historical travel paths, maximum capacity, and expected ports of arrival could also be extracted for other cruise ships in the region.
    
\end{itemize}	

\section{Discussion}
Ships represent a source of epidemic risk which might be particularly relevant in port cities and island nations or for diseases with long incubation periods. 
Our analysis of data from Madagascar and the Canary Islands illustrates that island shipping traffic has a broad reach (in these cases, reaching three or more continents within a month) and may potentially carry high volumes of passengers. Even a landlocked city such as Wuhan saw a considerable volume of river traffic in the two-week period prior to quarantine. 

AIS data makes it possible to visualize individual ship trajectories, along with additional ship information such as vessel type and size. It is simple to filter out, for instance, passenger ships that might have stopped in cities with high risk, as well as to account for their expected arrival times in future ports. The ability to acquire such information in real time might be of critical relevance for operational responses to disease outbreaks.

The approach described in this communication represents a first step towards estimating the potential role of vessel traffic in global disease spread. We assert that ship origin-destination matrices can offer insights into travel routes which may be overlooked when focusing on flight path data alone, and should be considered when modeling the spread of disease in port cities and island nations. The availability of real-time and historical data from commercial providers makes analysis of AIS data feasible.
Vessel trajectory data can be enhanced with other sources of data on the number of passengers and crew aboard tourist or commercial ships. Future work should explore how adding shipping traffic to existing flight network data changes the outcome of disease simulations and risk estimates. 

\section*{Acknowledgements}
We thank Daniela Perrotta for her contributions to the analysis of the Madagascar data. We thank the UN Global Platform for access to its AIS database  \cite{un_statistics_division_global_2020}. United Nations Global Pulse is supported by the Governments of the Netherlands, Sweden, and Germany and by the William and Flora Hewlett Foundation.

\bibliographystyle{ieeetr}          	
\bibliography{bibliography}

\newpage~\newpage
\section*{Appendix: Data and Methods}
\subsection*{AIS data}
Popular AIS data providers include ExactEarth, FleetMon, MarineTraffic, OrbComm, VesselFinder, and VesselTracker \cite{exactearth_ais_2020, fleetmon_live_2020, marinetraffic_global_2020, orbcomm_iot_2020,  vesselfinder_free_2020, vesseltracker_terrestrial_2020}. The frequency of AIS readings varies, with some providers offering better resolution than others; in our experience, we have found that this resolution can range from approximately one reading every two minutes to one reading per hour. Furthermore, providers differ in their access to land- or satellite-based receivers; typically, land-based receivers detect ships within 15 - 20 (and up to 60) nautical miles away \cite{all_about_ais_faqs_2020}, whereas satellite-based receivers are used to monitor ships further out to sea. 

\subsection*{Selecting AIS data for OD matrices}
\begin{figure}[b]
    \centering
    \includegraphics[width=2.75in]{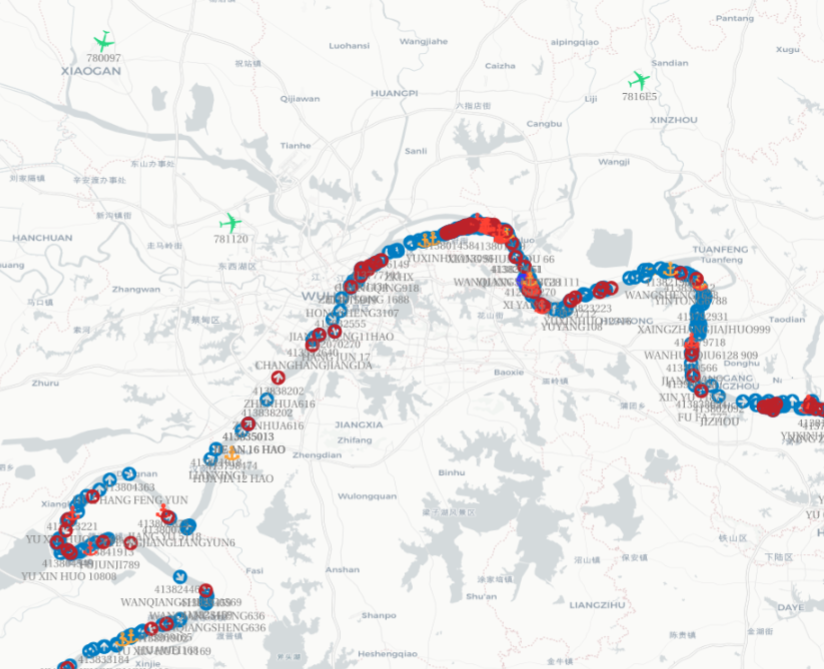}
    \includegraphics[width=2.75in]{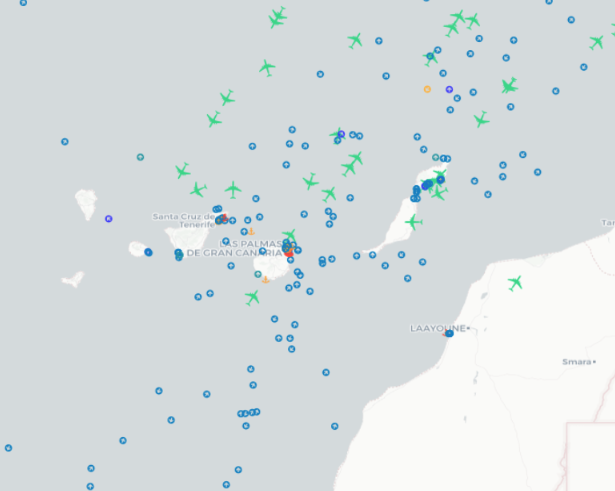}
    \caption{Snapshots of real-time plane and flight traffic from Wuhan (above) and the Canary islands (below) in February 2020. \textit{Source: UN Global Platform.}}
    \label{fig:data_snap}
\end{figure}

\begin{figure}[b]
    \centering
    \includegraphics[width=3in]{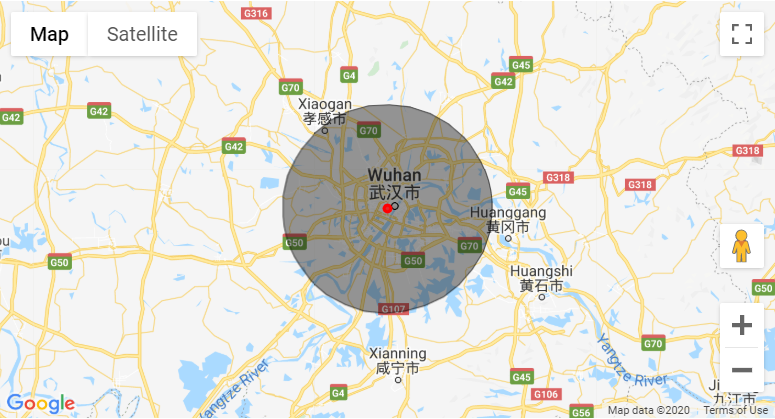}
    \includegraphics[width=3in]{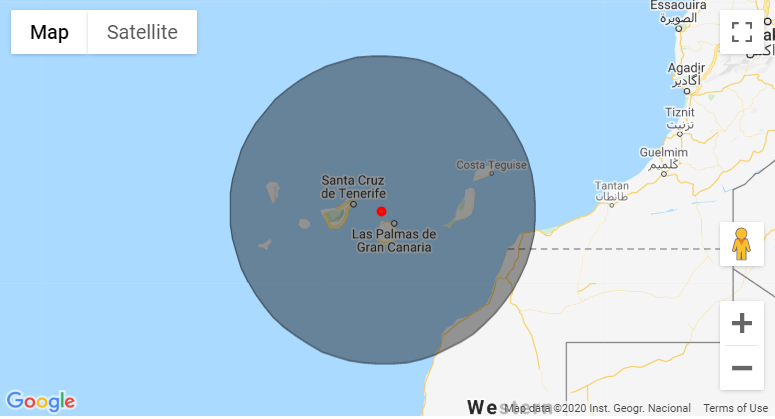}
    \caption{The radii used to select ships visiting Wuhan (above) and the Canary Islands (below). \textit{Basemap source: Google Maps.}}
    \label{fig:radius}
\end{figure}

We first queried the AIS database to find the Maritime Mobile Service Identities (MMSI) of all ships which had passed within 50km of Wuhan, or 300km of the Canary Islands, at a speed of less than 2 knots during our period of interest (8 - 21 January 2020). Examples of observed traffic in the areas of interest are shown in Figure \ref{fig:data_snap}, while the regions used for the query are shown in Figure \ref{fig:radius}. Given the resulting list of ship MMSIs, we then extracted all trajectory points for these ships during the period from 8 - 21 January, again retaining only points for which the ships recorded a speed of less than 2 knots.

\subsection*{Filtering AIS data for OD matrices}
We reduced the size of the AIS dataset using several criteria. First, we removed ships with invalid MMSI, dropping MMSI which were less than 9 digits in length and MMSI which were associated with multiple different vessel types (a sign that more than one ship might be using the same MMSI). Next, we removed duplicate readings; readings after which ships had moved less than 0.001 degrees in the vertical and horizontal directions from the previous reading; and ships whose position jumped at a rate of more than 1.7 degrees per hour (approximately 100 knots) in latitude or longitude   between subsequent position reports (another sign that more than one ship might be using the same MMSI). 

\subsection*{Building OD matrices from AIS data}
Using the list of ports supplied by the World Port Index \cite{national_geospatial-intelligence_agency_nautical_2019}, we next restricted our dataset to position readings which fell within approximately 50km of a port, and associated each of the remaining points with its nearest port. We then retrieved the port associated with the ship's previous location reading, dropping any connections between a port and itself. Finally, we aggregated the port-to-port connections of individual ships to yield an origin-destination (OD) matrix.

\end{document}